\documentclass[12pt,fleqn]{article}
\pdfoutput=1
\usepackage{verbatim,cite,amssymb,amsmath,graphicx,amsthm,subfigure,graphpap}
\numberwithin{equation}{section}

\theoremstyle{definition}

\theoremstyle{definition}

\renewcommand{\P}{{\cal P}}
\renewcommand{\L}{\Lambda}
\renewcommand{\H}{{\cal H}}
\newcommand{\C}{{\mathbb{C}}}

\newcommand{\<}{\langle}
\newcommand{\be}{\begin{equation}}
\renewcommand{\>}{\rangle}

\begin{document}

\title{Time Asymmetry in Quantum Physics - II. Experimental
  Demonstration\\ Using a Single Ion} \author{ A.~Bohm$^{\star}$,
  H.~Kaldass$^{\dagger}$, S.~Komy$^{\ddagger}$,
  W.P.~Schleich$^{\S}$\\[12pt]
  $^{\star}$\textit{\footnotesize CCQS, Physics Department,
    University of Texas,
    Austin, Texas 78712}\\
  $^{\dagger}$\textit{\footnotesize Arab Academy of Science and
    Technology, El-Horia, Heliopolis, Cairo, Egypt}\\
  $^{\ddagger}$\textit{\footnotesize Mathematics Department, Helwan
    University,
    Cairo, Egypt}\\
  $^{\S}$\textit{\footnotesize Institut f\"ur Quantenphysik,
    Universitat Ulm, Germany}\\
  {\footnotesize Emails: bohm@physics.utexas.edu,
    hani@ifh.de, sol\_komy@yahoo.com, Wolfgang.Schleich@uni-ulm.de}}
\date{}
\maketitle

\begin{abstract}
  Quantum physics involves an ensemble of quantum systems, usually one
  thinks of a large ensemble of identical quantum systems at one
  single time. In single ion experiments one has a single quantum
  system at an ensemble of different times. This provides the means of
  demonstrating the beginning of time of a semigroup evolution for a
  decaying state.
\end{abstract}

\section{Introduction -- states and observables}

In part~I of this series of papers we have discussed how time
asymmetry arose in quantum mechanics. Here we explain the meaning of
the time \mbox{$t=0$} of preparation of a quantum state. In the present paper
we describe an experiment which verifies our concept. It is based on
the idea of quantum jumps for a single ion. 

In essentially all discussions of the foundation of quantum theory,
one distinguishes between {\em states} and {\em observables}. States,
described by a state vector $\phi^+$ (or a density operator $W$)
fulfill the Schr\"odinger equation
\begin{equation}
\label{se}
i\hbar\frac{d}{d t}\phi^+(t) = H \phi^+(t)\,.
\end{equation}
The state represents the preparation apparatus of an experiment. 

Observables, described by operators $A = A^\dagger$ or projection
operators $\L$ ($\L = \L^2$) or, in the simplest case, by
one-dimensional projections $\L=|\psi^-\>\<\psi^-|$, or vectors (up
to a phase factor) $\psi^-$, fulfill the Heisenberg equation
\begin{subequations}
\label{he}
\begin{equation}
\label{hea}
i\hbar\frac{d}{d t}\psi^-(t) = -H\psi^-(t)\,
\end{equation}
\begin{equation}
\label{heb}
i\hbar\frac{d\Lambda(t)}{dt} = -[H,\Lambda(t)]\,,
\end{equation}
\end{subequations}
where \eqref{heb} for an operator $\Lambda(t)$, is the more familiar
form of this dynamical equation.  The observable represents the
registration apparatus (detector) of an experiment.

The experimental quantities are the probabilities $\P_W(\L(t))$ of the
observable $\L(t)$ in the state $W$. Theoretically, $\P_W(\L)$ are
calculated as the Born probabilities
\begin{align}
\label{bp1}
{\P}_W(\L(t)) = \text{Tr}(\L(t)W) = \text{Tr}(\L W(t))
\end{align}
which for the special case of an observable $|\psi^-\>\<\psi^-| =
\Lambda$ in a pure state $|\phi^+\>\<\phi^+|=W$ is given by
\begin{align}
  \label{bp2} {\P}_{\phi^+}(\psi^-(t)) = \begin{array}[t]{c}
    |\<\psi^-(t)|\phi^+\>|^2\\ \text{Heisenberg picture}\end{array} =
  \begin{array}[t]{c} |\<\psi^-|\phi^+(t)\>|^2\\ \text{Schrodinger
      picture}\end{array}\,.
\end{align}
Experimentally the probabilities $\P_W(\L)$ are measured as ratios of
large detector counts, e.g.,
\begin{equation}
\P_W(\L(t)) = \frac{N(t)}{N}\,, 
\end{equation}
where $N(t)$ is the number of detector count at time $t$, and $N$ is
the total number of counts.

The solutions of the dynamical differential equations like \eqref{se}
and \eqref{he} have to be found under {\em boundary conditions}.  For
standard quantum mechanics, the boundary condition is given by the
Hilbert space axiom:

The solutions of \eqref{se} {\em and} of \eqref{he} are elements of
the (norm-complete) Hilbert space $\H$.
\begin{equation}
\label{sa}
\text{set of states} \{\phi^+(t)\} = \H = \text{set of
  observables}\{\psi^-(t)\}\,.
\end{equation}
(This means the energy wavefunctions $\psi^-(E) = \<^-E|\psi^-\>$,
$\phi^+(E) = \<^+E|\phi^+\>$, must be {\em Lebesgues} square
integrable.)

In paper I \cite{paper1}, it was argued that under the standard axiom
\eqref{sa} one could not have a consistent theory of resonance and
decay phenomena: If one wants to include states for which the
approximate lifetime-width relation of the Weisskopf-Wigner methods,
$(2.9)$ of \cite{paper1}, becomes an exact equality, $\tau =
\hbar/\Gamma$, in particular if one wants to have Gamow vectors
$\phi^G(t)$ with exponential time evolution, then one cannot sustain
the standard axiom \eqref{sa}.  Therefore, we chose a new boundary
condition for the old dynamical equations \eqref{se}, \eqref{he}, the
{\em Hardy space axiom}:
\begin{equation}
\label{hsa}
\begin{split}
  &\text{Set of prepared (in-) states} & \{\phi^+\} =
  \Phi_-\subset\H\subset\Phi_-^\times \\
  &\text{Set of detected (out-) observables} & \{\psi^-\} =
  \Phi_+\subset\H\subset\Phi_+^\times
\end{split}
\end{equation}
where $\Phi_\mp$ are the two different Hardy spaces of the same
Hilbert space $\H$, corresponding to the complex energy semiplanes
$\mathbb{C}_\mp$.  This means the energy wavefunctions $\psi^-(E) =
\<^-E|\psi^-\>$ of out-observables must be smooth functions that can
be analytically continued into the upper complex energy plane $\C_+$,
and the wavefunctions $\phi^+(E) = \<^+E|\phi^+\>$ of the in-state
must be smooth functions that can be analytically continued into the
lower complex energy plane $\C_-$.

Since the in-states are defined by the preparation apparatus (e.g.,
accelerator) and the out-observables are defined by the registration
apparatus (e.g., detector) of an experiment on micro-systems, there is
no need to identify $\{\phi^+(E)\} = \{\psi^-(E)\}$ and use the
standard axiom \eqref{sa} which in the mathematical description
identifies states and observables with the Hilbert space.

\section{Time asymmetry}
A consequence of the new Hardy space axiom is the time asymmetry $t\ge
t_0 = 0$ of the dynamical evolution given by the semigroup solutions
$(4.6)$ and $(4.7)$ of \cite{paper1}. In particular the Gamow state
$\phi^G$ fulfilling $H^\times\phi^G = (E_R-i\Gamma/2)\phi^G$, evolves
in time according to $(4.6)$ of \cite{paper1} as
\begin{equation}
\label{gs}
\phi^G(t) = e^{-iH^\times t}\phi^G = e^{-iE_Rt}e^{-(\Gamma/2)
  t}\phi^G\,,\quad t\ge 0\,.
\end{equation}
and exists only for $t\ge t_0 = 0$. 

Since from the Hilbert space solution of the dynamical equations
\eqref{se}, \eqref{he}, we are used to time symmetric solutions,
$-\infty < t < +\infty$, such a time asymmetry may appear surprising.
But, from other areas of physics, we are quite familiar with
time-asymmetric solutions of time symmetric dynamical equations due to
time-asymmetric boundary conditions, e.g., the retarded and the
advanced solutions of electrodynamics.  The Maxwell equations, like
the equations \eqref{se} and \eqref{he}, are symmetrical in time.  The
additional axiom which chooses the retarded potential over the
advanced potential is called the ``radiation arrow of time''; it is
formulated as a boundary condition for the Maxwell equations, which
excludes the strictly incoming field in every space-time region
(Sommerfeld radiation condition). All fields possess advanced sources:
Radiation must be emitted first by a source before it can be detected
by a receiver.

A similar ``arrow of time'' exist for quantum physics: A state must be
prepared first by a preparation apparatus at $t_0=0$ before at a later
time $t > t_0$ an observable can be measured or registered by a
registration apparatus. In order to formulate this quantum mechanical
arrow of time we speak of states and observables, and we make a
distinction between the states $\{\phi^+\}$ and the observables
$\{\psi^-\}$. In quantum mechanics in general one always {\em speaks}
of states and observables, but then using the standard axiom
\eqref{sa} one identifies them in the mathematical description.
However, as discussed in \cite{paper1}, if we want to derive for the
pole term of the $S$-matrix element $(\psi^-,\phi^+)$ in $(3.8)$ of
\cite{paper1}, the relation $\Gamma = \hbar/\tau$ between the
Lorentzian width $\Gamma$ and the exponential lifetime $\tau$, then we
are forced to distinguish also mathematically between states and
observables and use the new axiom \eqref{hsa}. This in turn leads to
the semigroup evolution which introduces a ``beginning of time'', the
semigroup time $t_0 = 0$.

\section{Demonstrating the quantal beginning of time}
The question in quantum physics then is: What is the meaning of a
``beginning of time $t_0$'' ?  How does one observe it and why have we
not been more aware of its existence ?

Most experiments in quantum physics deal with large ensembles of
quantum systems (elementary particles, or atoms, or ions) which are
prepared at a variety of times. That means individual atoms of an
ensemble are prepared at many different times and there is no way to
distinguish which has been prepared at a time, say $t_0^{(1)}$ and
which at a time $t_0^{(2)}$ (by the clocks in the lab). If, however,
one could work with {\em single quantum systems and identify the
  preparation time of a single quantum system}, then one could
consider the quantum mechanical state, say $\phi^G$, whose ensemble is
the collections of the single quantum systems prepared at the
``ensemble of times'' $t_0^{(1)}$, $t_0^{(2)}$, $t_0^{(3)}$, $\cdots$

Such experiments on the excited states of an individual quantum system
have been performed in experiments with single laser cooled $Ba^+$
ions \cite{nagourney, sauter}, $Hg^+$ ions \cite{bergquist} and $In^+$
ions \cite{peik} using Dehmelt's idea of shelving the single ion on a
metastable level \cite{dehmelt}.  Experiments of this kind require an
atom with two excited states both of which are radiatively coupled to
the same ground state but have vastly different transition rates.  We
will discuss here in detail the experiment in \cite{peik} with an
$In^+$ ion.

\begin{figure}
\begin{center}
\begin{picture}(300,200)(0,0)
\put(10,0){\line(1,0){60}}
\put(40,0){\line(0,1){80}}
\put(40,180){\line(0,-1){80}}
\put(10,180){\line(1,0){60}}
\put(80,140){\line(1,0){50}}
\put(130,115){\line(1,0){50}}
\put(180,90){\line(1,0){50}}
\put(155,115){\vector(-1,-1){114}}
\put(155,115){\vector(2,-1){50}}
\put(205,90){\vector(-3,-2){135}}
\put(10,5){\makebox(0,0)[b]{$5s^2\ ^1S_0$}}
\put(10,185){\makebox(0,0)[b]{$5s\,5p\ ^1P_1$}}
\put(190,110){\makebox(0,0){$A_2$}}
\put(85,65){\makebox(0,0){$A_1$}}
\put(155,40){\makebox(0,0){$A_0$}}
\put(160,125){\makebox(0,0)[c]{$\,^3P_1$}}
\put(210,100){\makebox(0,0)[c]{$\,^3P_0$}}
\put(100,85){\makebox(0,0)[b]{$231\,nm$}}
\put(190,60){\makebox(0,0)[b]{$237\,nm$}}
\put(80,150){\makebox(0,0){$5s\,5p\ ^3P_2$}}
\put(40,90){\makebox(0,0){$159\,nm$}}
\put(70,120){\makebox(0,0){$218\,nm$}}
\end{picture}
\end{center}
\caption{Simplified energy-level scheme of the $In^+$ ion. \cite{peik}}
\label{fig1}
\end{figure}
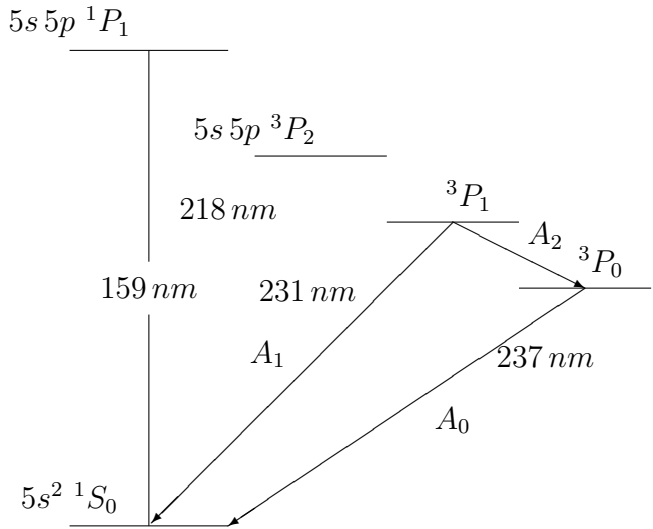

For the $In^+$ these two levels are the $\,^3P_1$ and $\,^3P_0$
levels, Fig.(\ref{fig1}) \cite{peik}.
We denote the  transitions involved by $A_1$, by $A_2$ and $A_0$.
One has the following resonance scattering and decay processes
\begin{equation}
\label{dp}
\begin{array}{cccccc}
  \gamma +\ ^1S_0 & \begin{array}{l} 
    \underleftarrow{ \begin{array}{l}\text{fluorescence} \\
        \text{transition $A_1$\phantom{ition}}\end{array} } \\
    \overrightarrow{ \begin{array}{l}\text{laser driven} \\ \text{fast
          transition $A_1$\ }\end{array} }
  \end{array} & \,^3P_1 & \begin{array}[t]{c}
    \xrightarrow[\quad\qquad\quad]{}\\ A_2\end{array}  &
  \text{\begin{picture}(0,0)(0,-5)
      \put(0,0){\makebox(0,0)[cl]{$\gamma^\prime +\, ^3P_0$}}
      \put(35,-5){\line(0,-1){20}}
      \put(35,-25){\vector(1,0){20}}
      \put(55,-25){\makebox(0,0)[cl]{$\,^1S_0 + \gamma^{\prime\prime}$}}
      \put(35,-33){\makebox(0,0){$A_0$ slow}}
\end{picture}} & 
\end{array}
\end{equation}
We focus our attention on the excited (metastable) level $\,^3P_0$
which is the state we describe by the Gamow vector $\phi^G =
|z_R,J^\pi,\cdots\> = |''3P_0''\>$.

The Gamow vector is an eigenket of the total, interaction
incorporating Hamiltonian $H$ with complex energy eigenvalue $z_R =
E_R-i\Gamma_R/2$ where $\Gamma_R = \frac{\hbar}{\tau}$, and $\tau =
\tau(\,^3P_0)\approx 0.14\,s$.  We use here the standard spectroscopic
notation, e.g., $\,^3P_0 =\, ^{2s+1}(L)_J$, with $s=1$ and $J=0$ for
the excited energy levels, but $H$ is the exact Hamiltonian which
includes spin orbit {\em and} hyperfine interaction, and $\phi^G$ is
an eigenstate of $H$ and of total angular momentum-parity, $J^\pi$.
Due to the hyperfine interaction, the state of the $\,^3P_0$ level
includes also a superposition with a very small $J=1$
contribution\footnote{due to the magnetic dipole component of the
  hyperfine interaction \cite{garstang}} which in turn is a
superposition of a small amount
of triplet and  singlet states: \\
($|''\,^3P_0''\> = |\,^3P_0\> + \alpha|\,^3P_1\> + \beta|\,^1P_1\>$).
Due to this $J=1$ component there is a ``slow'' transition $A_0$.

The transition $\gamma +\, ^1S_0\rightarrow\, ^3P_1$ is laser driven
and the intensity of the fluorescence transition $A_1$:
$\,^3P_1\rightarrow\,^1S_0 + \gamma$ (with a lifetime
$\tau(\,^3P_1)\approx 4\times 10^{-7}\,s$) is monitored. Occasionally
$\,^3P_1$ makes the magnetic dipole transition into $\,^3P_0$ with a
branching ratio $10^{-8}$ (there can also be a laser induced
transition from $\,^1S_0$ to $\,^3P_0$). This is the metastable state
$|''\,^3P_0''\>\equiv \phi^G$ in which the ion will be shelved for a
long time ($\tau(\,^3P_0) \approx 0.14\,s$).

The experiment is done on a {\em single} $In^+$ and while it is
shelved it can not participate in the back and forth transitions
\begin{equation}
\label{mt}
\,^1S_0 + \gamma \overleftarrow{\overrightarrow}\,^3P_1
\end{equation}
so the shelve time is observed as a dark period of the fluorescence
transition $A_1$: $\,^3P_1\rightarrow\,^1S_0 + \gamma$.

The state of $In^+$ could be either the metastable state
$|''\,^3P_0''\> = |\phi^G(t)\>$ evolving in time by \eqref{gs}, or it
could go through many $(10^6)$ back-and-forth transitions \eqref{mt}
or it could evolve through superpositions of these three states.

\begin{figure}[h]
  \centering
  \includegraphics[width=\textwidth]{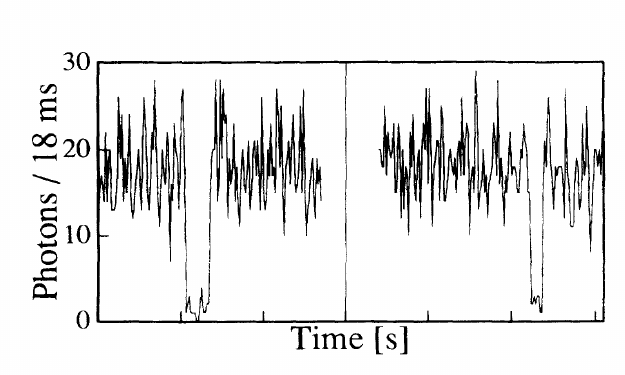}
  \caption{Two examples of observed quantum jumps resulting in dark
    periods~\cite{peik}. The sudden drop in fluorescence defines the
    beginning of time in $i$-th individual quantum state
    $''^{3}P_{0}''$. The duration of the dark period $\Delta t^{(i)}$ is
  the individual lifetime of this $''^{3}P_{0}''$. The average lifetime
  $\tau$ is the weighted average of these $\Delta t^{(i)}$.}
  \label{fig2}
\end{figure}

\begin{figure}[h]
  \centering
  \includegraphics[width=\textwidth]{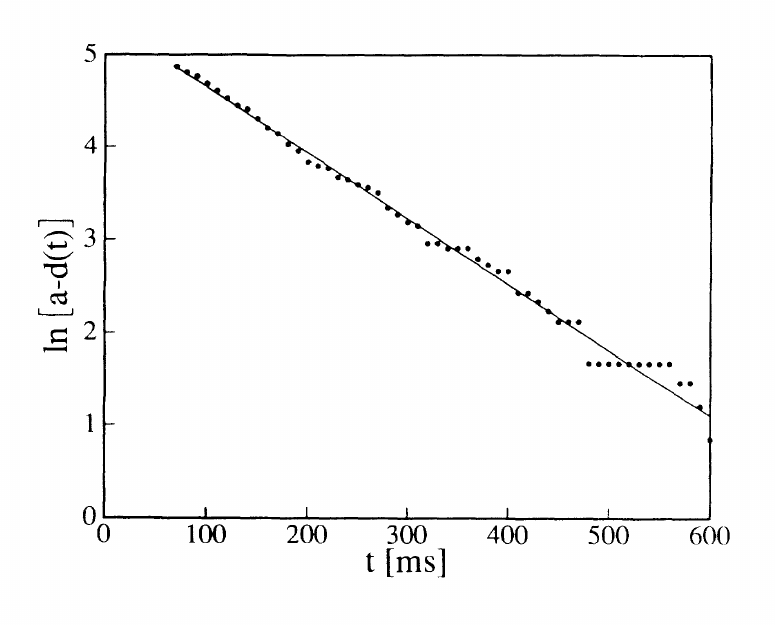}
  \caption{The fit of the number of dark periods shorter than $t$ to
    the experimental law~(\ref{tau}) provides an estimate of the
    average lifetime of the $^{3}P_{0}$ level.}
  \label{fig3}
\end{figure}

What one observes (Fig.~\ref{fig2}, \cite{peik}) is a sudden onset of
periods of no fluorescence at a time $t_0^{(1)}$ and a sudden return
of the original fluorescence intensity at a later time $t^{(1)}$.
These onsets of dark periods followed by the return of the
fluorescence repeat themselves (about $150$ times in the experiment
\cite{peik}): at $t_0^{(1)},t_0^{(2)},\cdots,t_0^{(150)}$ one sees
sudden jumps of the fluorescence intensity to $0$, followed by a
sudden return of the fluorescence radiation at the later times
$t^{(1)},t^{(2)},\cdots,t^{(150)}$.

The only reasonable interpretation of this is that during the dark
periods the single $In^+$ ion is shelved on the level $''\,^3P_0''$.
While $In^+$ is ``shelved'' on the metastable level $\,^3P_0$, it
cannot participate in the fluorescence transitions
$\,^3P_1\overrightarrow{\overleftarrow}\, ^1S_0 + \gamma$ and one
obtains dark periods (``shelf level dwell period'') of various
durations
\begin{equation}
\label{vd}
\Delta t^{(i)} = t^{(i)} - t_0^{(i)}\,,\quad \Delta t^{(i)} >
0\,,\quad t^{(i)} > t_0^{(i)}\,. 
\end{equation}
Two of these dark periods for the fluorescence transition are shown in
Fig.~\ref{fig2}. A dark period means that the system is on
the metastable level $\,^3P_0$, which is described by $\phi^G =
|''\,^3P_0''\>$; this means that at every of the times $t_0^{(i)}$ an
individual $\,^3P_0$ is prepared, it ``lives'' for the duration
$\Delta t^{(i)}$ and decays at the time $t^{(i)}$.

The state vector $\phi^G(t)$ for the meta-stable $''\,^3P_0''$-state
therefore represents an ensemble of individual $\,^3P_0$-levels of
$In^+$; the $i$-th member of this ensemble is created (or prepared) at
a definite time $t_0^{(i)}$ which is the time $t=0$ in the life of
this $i$-th member. This is also the time $t=0$ of the state
$\phi^G(t)$.

A quantum mechanical state vector $\phi^+$, or $\phi^G$, or state
operator $W$, represents an ensemble (a large number) of individual
micro-systems. Usually one thinks of an ensemble of microsystems as of
collections of many particles that are present at one particular point
of time (in our live, or by the laboratory clocks); in this experiment
one has one single ion in the $\,^3P_0$ level at a collection of many
preparation times $\{t_0^{(i)}\}$. The preparation time for this
ensemble of single shelved $In^+$ is the ensemble of times
$\{t_0^{(i)}\}$.  This ensemble of times $\{t_0^{(i)}\}$ is the
preparation time $t_0 = 0$ of the state operator
$|\phi^G(t)\>\<\phi^G(t)|$. The state vector $\phi^G(t)$ which
represents the ensemble of $In^+$ in the $\,^3P_0$ level has thus a
time evolution $0\le t<\infty$, and the semigroup time $t=0$ is the
ensemble of times $\{t_0^{(i)}\}$ where the times $t_0^{(i)}$ are the
onset times of dark periods, i.e., the times at which the $\,^3P_1$
level stops participating in the fluorescence transitions \eqref{mt}
and prepares the state $\phi^G = |''\,^3P_0''\>$.

Since the lifetime $\,^3P_1$ is $10^{-6}\,s$, and the lifetime of
$''\,^3P_0''$ is $10^{-1}\,s$, the onset time of the dark periods
$t_0^{(i)}$ are determined with very high accuracy. The same applies
to the time $t^{(i)}$, since one observes a {\em sudden} return of
fluorescence.  This is the time at which the $i$-th
$''\,^3P_0''$-level has decayed into $\,^1S_0 + \gamma$, and the
processes \eqref{mt} can resume.  The duration of a dark period
$\Delta t^{(i)} = t^{(i)}-t_0^{(i)}$ is thus the individual lifetime
of the $i$-th $\,^3P_0$ level, and it is defined to very high
accuracy. Two of these individual lifetimes $\Delta t^{(i)}$ of the
$\,^3P_0$ levels are shown by the duration of the dark periods in
Fig.~\ref{fig3}, \cite{peik}.

Summarizing, the interpretation of our observation is the following:
Since the single $In^+$ can either make the fluorescence monitored
transitions \eqref{mt} or be in the quasistable state $|''\,^3P_0''\>$
the time interval $\Delta t^{(i)} = t^{(i)}-t_0^{(i)}$ is the time
which the $i$-th single $In^+$ at the $''\,^3P_0''$ level of $In^+$
has ``lived'', from its preparation at $t_0^{(i)}$ to its decay into
$\,^1S_0 + \gamma$ at $t^{(i)}$.

\section{The exponential of the average lifetime}
The ensemble of the $''\,^3P_0''$ levels is described by the state
vector $\phi^G(t)$. The preparation time $t_0$ of the state
$\phi^G(t)$ is the ensemble of times $\{t_0^{(i)}\}$,
\begin{equation}
\label{new1.12}
\,\quad\qquad \quad t_0 \Leftrightarrow \{t_0^{(i)}\}\,.
\end{equation}
This time $t_0 > -\infty$ is the time zero in the existence of each
individual $''\,^3P_0''$-$In^+$-ion state. We choose it thus as the
time $t = 0$ of the time evolution semigroups
\begin{equation}
\label{new1.13-}
U_-^\times(t) = \{ e^{-iH^\times t}\,|\,0\le t < \infty \}
\end{equation}
for the Gamow state vector $\phi^G(t)$ of \eqref{gs}. And similarly
one defines the semigroup for any state $\phi^+\in\Phi_-$ in $(4.6)$
of \cite{paper1}.

Therewith we have identified the experimental definition of the
semigroup time $t_0 = 0$, of the state $\phi^G(t)$: it is the ensemble
of onset times of the dark periods $\{t_0^{(i)}\}$ when the $In^+$ is
shelved on the $\,^3P_0$ level, and where it remains for an ensemble
of individual lifetimes $\Delta t^{(i)}$.

The quantities predicted by quantum theory are the Born probabilities
and averages (expectation values) not the properties of individual
quantum systems like the $\Delta t^{(i)}$. The Born probability for
the Gamow state \eqref{gs} (survival probability) is given by the
exponential law
\begin{equation}
\label{new1.14m}
\P(\,^3P_0(t)) \equiv \P_D(t) = e^{-\Gamma t/\hbar} =
e^{-t/\tau}\,,\qquad t > 0\,.
\end{equation}
These probabilities are measured by the counting ratios:\footnote{the
  sign $\simeq$ in \eqref{new1.14} and below indicates that this is an
  equality between theoretical quantities on the left hand side and
  experimental quantities on the right hand side (which would become
  an exact equality in the unrealistic case of continuously infinite
  events).}
\begin{equation}
\label{new1.14}
e^{-t/\tau} = \P_D(t) \simeq \frac{N_D(t:\Delta t^{(i)} > t)}{N_D} =
\text{counting ratio}
\end{equation}
where 
\begin{equation*}
N_D(t:\Delta t^{(i)} > t) = \text{number of dark periods of
  duration}\quad \Delta t^{(i)} > t\,, 
\end{equation*}
and $N_D$ is the total number of dark period events that are included.
We call
\begin{equation*}
  N(t:\Delta t^{(i)} < t ) = \text{number of dark periods of
    duration}\quad \Delta t^{(i)} \le t\,.
\end{equation*}
Intuitively, $N_D(t:\Delta t^{(i)} > t)$ is the number of $\,^3P_0$
levels that live longer than $\Delta t^{(i)}$, and $N(t:\Delta t^{(i)}
< t )$ is the number of $\,^3P_0$ levels that have already decayed
into $\,^1S_0 + \gamma$.

It is clear that for every time $t\ge 0$
\begin{equation}
\label{new1.15}
N(t:\Delta t^{(i)}\le t) + N_D(t:\Delta t^{(i)}>t ) = N_D
\end{equation}
where $N_D(\approx 150)$ is the total number of dark periods that have
been observed.  To check the exponential law \eqref{new1.14} and to
obtain an estimate of the average lifetime $\tau = \Gamma/\hbar$, the
number of dark periods with duration shorter than $t$, $N(t:\Delta
t^{(i)}\le t)\equiv d(t)$ is considered. Only dark periods of duration
larger than $t_s = 70\,ms$ could be identified in the experiment
\cite{peik}. Since from \eqref{new1.15} with \eqref{new1.14} follows
\begin{multline}
N_D(t:\Delta t^{(i)} > t)  = N_D e^{-t_s/\tau}e^{-(t-t_s)/\tau} = N_D
- N(t:\Delta t^{(i)} < t ) \\
 = a' e^{-(t-t_s)/\tau} = a - d(t)\,,
\end{multline}
the quantities $a-d(t)$ are plotted versus time in a logarithmic plot,
Fig.~\ref{fig3}.  The straight line confirms the exponential decay law
and from the slope of the straight line one obtains the lifetime of
the $\,^3P_0$-level as $\tau(\,^3P_0) \approx 0.14\,s$.

Since the counting ratios in \eqref{new1.14} are according to
Fig.~\ref{fig3} in agreement with the exponential law, the average of
the lifetime is given by $\tau$:
\begin{equation}
\label{tau}
\int_0^\infty dt\, \P_D(t) = \int_0^\infty dt\,e^{-t/\tau} = \tau\,.
\end{equation}
Since according to \eqref{new1.14} $\P_D(t)$ is observed as the
counting ratio\\
$N_D(t:\Delta t^{(i)}>0)/N_D$, the lifetime is measured as the
weighted average of the dark periods $\Delta t^{(i)}$:
\begin{equation}
\label{sld}
\tau \simeq \sum_i \Delta t^{(i)}\frac{N_D(t:\Delta t^{(i)} > t)}{N_D}\,.
\end{equation}
The ensemble of quasistable quantum objects in the state $\phi^G$ is
an ensemble of individual $''\,^3P_0''$s each of which is prepared at
the collection of times $t_0^{(i)}$ and decays at the collections of
times $t^{(i)}$ and lives for a duration of time $\Delta t^{(i)} =
t^{(i)} - t_0^{(i)}$.  Not the values $\Delta t^{(i)}$ are
characteristic quantities of the Gamow state but the weighted average
\eqref{sld} which corresponds to the theoretical lifetime $\tau$ of
the Gamow state in \eqref{gs} with $\tau = \Gamma/\hbar$.

The lifetime $\tau$ is one of the characteristics of the Gamow state
$(3.12)$ of \cite{paper1}.  It is defined by the exponential law
\eqref{new1.14m} which follows from \eqref{gs}. It is given by the
$S$-matrix pole position $z_R = E_R-i\Gamma/2$ which defines a complex
energy eigenvector $(3.14)$ of \cite{paper1} which describes a
resonance state of Breit-Wigner width $\Gamma$. And the lifetime of
the decaying state is related to the width of the resonance by the
exact lifetime-width relation $\tau = \frac{\hbar}{\Gamma}$.

\section{Conclusion}
The time asymmetry of quantum theory which was shown in \cite{paper1}
to be a byproduct of a unified theory of resonance scattering and
decay phenomena, distinguishes a particular time $t_0$, the time $t =
0$ of a semigroup time evolution. This is the time at which the state
(e.g., decaying state) has been prepared and the registration of the
observable (e.g., decay products) can begin.

This semigroup $t=0$ is measured as the collection of times
$\{t_0^{(i)}\}$ at which the individual $\,^3P_0$-levels of a single
$In^+$ were prepared.  These times of preparation of $\,^3P_0$ levels
are documented by the sudden onset times $t_0^{(i)}$ of the dark
periods for the single $In^+$ ion. Each dark period also has its
individual end $t^{(i)}$ (documented by the end of the dark period).
Thus each dark period has its individual lifetime $\Delta t^{(i)} =
t^{(i)}-t_0^{(i)}$.  The quantum mechanical state $\phi^G$ describes
the ensemble of individual microsystems created under the same
condition at an ensemble of different times $\{t_0^{(i)}\}$. This
ensemble of preparation times is the preparation time $t_0=0$ of the
state $\phi^G$: $t_0 \leftrightarrow \{t_0^{(i)}\}$.

In traditional quantum mechanics the state would be asymptotically
prepared for $t\rightarrow -\infty$ as a state of the
$(\,^1S_0,\gamma)$-system and evolve in time through a superpositions
of vectors representing $(\,^1S_0,\gamma)$, $\,^3P_1$,
$(\gamma',\,^3P_0)$, $(\,^1S_0,\gamma'')$,$\cdots$.  Using the Hilbert
space axiom \eqref{sa}, the probability for $\,^3P_0$ would be
different from zero -- at least infinitesimally -- at any time $t$
between $-\infty < t < \infty$. It could be significantly different
from zero at a particular time $t_0$, but there could not be sudden
jumps as used for the interpretation of the sudden onset of dark
periods at $t_0^{(i)}$.

On the other hand the existence of these quantum jumps is an
experimental confirmation of the semigroup time evolution. The new
Hardy space hypothesis \eqref{hsa} which was conjectured in order to
obtain a consistent theory that unifies Breit-Wigner resonances and
exponentially decaying Gamow states also led to a semigroup evolution
\cite{paper1} and therewith the semigroup time $t_0 = 0$. This time
$t_0$ represents the ensemble of preparation times $\{t_0^{(i)}\}$ for
the metastable level $\,^3P_0$ which is described by the Gamow state
vector $\phi^G = |''\,^3P_0''\>$.

Quantum theory describes ensembles of (large) numbers of quantum
systems. In the usual experiments this means or at least includes a
large number of micro-systems at any given time.  In the experiments
with single ions, it involves only one single ion state which is
prepared at a large number of different times $t_0^{(i)}$ under
identical conditions. This exposes the time of preparations for the
state of the single quantum system. What is remarkable about these
marvelous experiments \cite{sauter,bergquist,peik} is that they can
measure quantities that the theory cannot predict, like the individual
lifetimes of a single excited ion level.

\section*{Acknowledgment}
This work is part of a collaboration sponsored by the US National
Science Foundation Award No.  OISE-0421936.

\end{document}